\documentclass[12pt]{iopart}

\usepackage{hyperref}
\usepackage{graphicx}
\usepackage{amsmath}
\usepackage{amssymb}
\usepackage[dvipsnames]{xcolor}
\usepackage{mathtools}

\usepackage[normalem]{ulem}

\usepackage[nolist,nohyperlinks]{acronym}
\acrodef{AGN}{active galactic nucleus}
\acrodefplural{AGN}{active galactic nuclei}
\acrodef{CMB}{cosmic microwave background}
\acrodef{CTA}{Cherenkov Telescope Array}
\acrodef{DSR}{deformed (or doubly) special relativity}
\acrodef{EBL}{extragalactic background light}
\acrodef{GR}{general relativity}
\acrodef{GRB}{gamma-ray burst}
\acrodef{HE}{high-energy}
\acrodef{IACT}{imaging air-Cherenkov telescope}
\acrodef{ICS}{inverse Compton scattering}
\acrodef{IGMF}{intergalactic magnetic field}
\acrodef{LHAASO}{Large High Altitude Air Shower Observatory}
\acrodef{LIV}{Lorentz invariance violation}
\acrodef{PD}{photon decay}
\acrodef{PP}{pair production}
\acrodef{QCD}{quantum chromodynamics}
\acrodef{QED}{quantum electrodynamics}
\acrodef{QG}{quantum gravity}
\acrodef{SM}{Standard Model}
\acrodef{SR}{special relativity}
\acrodef{SWGO}{Southern Wide-field Gamma-ray Observatory}
\acrodef{UHECR}{ultra-high-energy cosmic ray}
\acrodef{VC}{vacuum Cherekov}
\acrodef{VHE}{very-high-energy}

\allowdisplaybreaks

\begin{document}

\title[Simulating Electromagnetic Cascades with Lorentz Invariance Violation]{Simulating Electromagnetic Cascades with Lorentz Invariance Violation}

\author{Andrey Saveliev$^{1,2}$, Rafael {Alves Batista}$^{3,4,5}$}
$^{1}$Immanuel Kant Baltic Federal University, Ul.~A.~Nevskogo 14, Kaliningrad, Russia \\
$^{2}$Lomonosov Moscow State University, GSP-1, Leninskiye Gory 1-52, Moscow, Russia \\
$^{3}$Instituto de F\'isica Te\'orica UAM-CSIC, Universidad Aut\'onoma de Madrid, C/ Nicol\'as Cabrera 13-15, 28049 Madrid, Spain\\
$^{4}$Departamento de F\'isica Te\'orica, Universidad Aut\'onoma de Madrid, M-15, 28049 Madrid, Spain\\
$^{5}$Sorbonne Universit\'e, CNRS, UMR 7095, Institut d'Astrophysique de Paris, 98 bis bd Arago, 75014 Paris, France
\ead{andrey.saveliev@desy.de, rafael.alves\_batista@iap.fr}

\begin{abstract}

\ac{LIV} is a phenomenon featuring in various quantum gravity models whereby Lorentz symmetry is broken at high energies, potentially impacting the behaviour of particles and their interactions. Here we investigate the phenomenology of \ac{LIV} within the context of gamma-ray--induced electromagnetic cascades. We conduct detailed numerical simulations to explore the expected manifestations of \ac{LIV} on gamma-ray fluxes, taking into account relevant effects such as pair production and inverse Compton scattering. Additionally, we consider processes forbidden in the Standard Model, namely vacuum Cherenkov emission and photon decay.
Our analysis reveals that these modifications result in distinct characteristics within the measured particle fluxes at Earth, which have the potential to be observed in high-energy gamma-ray observations.

\end{abstract}

\maketitle

\tableofcontents

\section{Introduction}

The \ac{SM} of particle physics is a highly successful theory that provides an elegant description of three out of the four known fundamental forces of nature, namely electromagnetic, weak, and strong interactions, leaving aside only gravity~\cite{Gaillard:1998ui, CottinghamSM}. A unified theory describing these four forces -- if it exists -- would completely transform our understanding of nature and is arguably one of the most important problems in fundamental physics.

The \ac{SM} has proven remarkably effective in explaining a wide array of physical phenomena. Nevertheless, several unresolved issues persist, stemming from both experimental observations and theoretical considerations. These include phenomena like dark matter, dark energy, neutrino masses, and the hierarchy problem~\cite{Golling:2016gvc}.

\Ac{QG} theories attempt to describe the four fundamental forces within a single theoretical framework~\cite{Addazi:2021xuf}. It generally strives to bridge the gap between the quantum field theory approach of the \ac{SM} and the differential geometry framework employed in \ac{GR}. This task presents a significant challenge due to the classical nature of GR, which inherently complicates its combination with any form of quantization. 

\ac{QG} is expected to manifest itself only at high energies, comparable to the Planck mass ($M_{\rm Pl} \simeq 1.22 \times 10^{28} \; \text{eV}$) -- otherwise it would probably have already been detected in particle accelerators, whose maximal attainable energy is more than a dozen orders of magnitude below $M_{\rm Pl}$. Over very long baselines, Planck-scale effects can be magnified and lead to a measurable signal, as is the case of high-energy particles propagating over cosmological distances~\cite{Addazi:2021xuf}, such as \acp{UHECR}~\cite{Mattingly:2005re, Bietenholz:2008ni, JCAP10-03-046} and very-high-energy gamma rays~\cite{Mattingly:2005re,Galaverni:2008yj, Martinez-Huerta:2020cut}.

There have been several works positing that \ac{QG} lead to a phenomenon known as 
\acl{LIV}~\cite{Colladay:1998fq,Alfaro:2004aa,Collins:2004bp,Sotiriou:2009bx,Reyes:2014wna}. Lorentz symmetry can be broken if at least one of the pillars upon which it rests is broken, namely the principle of relativity, which ensures the equivalence of the laws of physics in all inertial frames, or the constancy of the speed of light. The immediate phenomenological consequence of \ac{LIV} is, thus, a modification of the energy-momentum dispersion relation and spatio-temporal changes. Note that some \ac{QG} frameworks also feature modified dispersion relations whilst merely deforming Lorentz symmetry. This, for example, is the case for \ac{DSR}~\cite{Amelino-Camelia:2002cqb, Morais:2023amp}. 

While there are currently no compelling hints of \ac{LIV}, it could either be small or manifest itself at energies beyond current observational limits. Therefore, attempts to perform experimental searches for \ac{LIV} often employ an effective field theoretical framework, like \ac{SM} extensions~\cite{Colladay:1998fq}, whose operators directly translate into \emph{measurable} phenomenological observables such as energy-dependent time delays~\cite{Amelino-Camelia:1997ieq, Ellis:2005sjy, Ellis:2011ek, Ellis:2018lca, jacob2008a} and anomalous interaction thresholds~\cite{Jacobson:2002hd,Mattingly:2002ba,Bertolami:2003yi,Baccetti:2011us,JCAP10-03-046,Martinez-Huerta:2018hpt} (see~\cite{Addazi:2021xuf} for a comprehensive review).

In this work we focus on the expected signatures of \ac{LIV} on gamma rays from cosmological sources, considering their intergalactic propagation~\cite{Batista:2021rgm}. During their journey from an astrophysical object to Earth, sufficiently energetic photons from an astrophysical object like a blazar or a gamma-ray burst interact with radiation pervading the universe, such as the \ac{CMB} and the \ac{EBL}, producing an electron-positron pair. These pairs, in turn, interact with background photons via inverse Compton scattering, transferring a substantial fraction of their energy to the photons resulting from it. These photons, depending on their energy, can trigger these interactions yet again, until the high-energy photons no longer have sufficient energy to start the reaction. Therefore, a single primary photon of sufficiently high energy can result in copious amounts of secondaries, tertiaries, and higher-order gamma rays. This energy \emph{threshold} is modified if \ac{LIV} is acting~\cite{Vankov:2002gt,Jacobson:2002hd}, changing the interaction length for pair production and igniting or quenching the cascade.

In this work, we perform Monte Carlo simulations of the development of electromagnetic cascades, considering the modified dispersion relations that affect \ac{VHE} photons and electrons during their propagation. 
To this end, we begin in Sec.~\ref{sec:ECLIV} by introducing the formalism governing the propagation of electrons and positrons, emphasising modifications stemming from the modified dispersion relations.
Subsequently, in Sec.~\ref{sec:Simulations}, we outline the simulation setup for the electromagnetic cascades incorporating \ac{LIV} effects and present the corresponding simulation results. In Sec.~\ref{sec:Discussion} we discuss our findings and the implications for searches of \ac{LIV} using high-energy gamma-ray observations. Lastly, in Sec.~\ref{sec:Conclusions} we draw our conclusions and briefly address the issue of employing this strategy for \ac{LIV} studies.

\section{Particle Propagation with Modified Dispersion Relations} \label{sec:ECLIV}

The aforementioned dispersion relations relevant for electromagnetic cascades are in general given by \cite{Terzic:2021rlx}
\begin{eqnarray}
&E_{e}^{2} = p_{e}^{2} \left[ 1 + \frac{m_{e}^{2}}{p_{e}^{2}} + \sum_{j=0}^{N}\chi_{j}^{e} \left(\frac{p_{e}}{M_{\rm Pl}}\right)^{j} \right] \,, \label{EeDispFull} \\
&E_{\gamma}^{2} = k_{\gamma}^{2}\left[ 1 + \sum_{j=0}^{N}\chi_{j}^{\gamma} \left(\frac{k_{\gamma}}{M_{\rm Pl}}\right)^{j} \right] \label{EgammaDispFull}
\end{eqnarray}
for electrons/positrons and photons, respectively. Here, $m_{e}$ is the electron mass, while $p_{e}$ and $k_{\gamma}$ is the electron and photon momentum, respectively. On the other hand, $\chi_{j}^{\gamma}$ and $\chi_{j}^{e}$ are the corresponding \ac{LIV} parameters of the $j$th order. 

Usually one assumes that one of the terms of the sums (of order $n = 0, 1, 2, ... $) present in Eqs.~(\ref{EeDispFull}) and (\ref{EgammaDispFull}) is dominant, such that these equations may be reduced to
\begin{eqnarray}
&E_{e}^{2} = p_{e}^{2} \left[ 1 + \frac{m_{e}^{2}}{p_{e}^{2}} + \chi_{n}^{e} \left(\frac{p_{e}}{M_{\rm Pl}}\right)^{n} \right] \,, \label{EeDisp} \\
&E_{\gamma}^{2} = k_{\gamma}^{2}\left[ 1 + \chi_{n}^{\gamma}\left(\frac{k_{\gamma}}{M_{\rm Pl}}\right)^{n} \right]\,, \label{EgammaDisp} 
\end{eqnarray}
respectively. In case of $\chi_{n}^{i} < 0$ the corresponding \ac{LIV} scenario is also called subluminal, while for $\chi_{n}^{i} > 0$  it is denoted as superluminal. Evidently, $\chi_{n}^{i} = 0$ reduces to the usual case of \ac{SR}.

The consequences of these modifications are two-fold \cite{Jacobson:2002hd}. Firstly, the threshold of a given reaction is altered, leading to a modification in the associated propagation length, as described in Sec.~\ref{ssec:thresholds}. Secondly, the presence of \ac{LIV} enables new reactions that would otherwise not be possible within the confines of Lorentz invariance. Examples of such reactions include the spontaneous decay of photons into pairs or photons, as well as the occurrence of the vacuum Cherenkov effect for electrons, both described in Sec.~\ref{ssec:newProcesses}.

\subsection{Modified Thresholds for Electromagnetic Interactions}\label{ssec:thresholds}

Here, we primarily consider the threshold modification. In order to calculate the threshold of a given reaction, one has to solve the equation 
\begin{equation} \label{ThrCond}
\max_{0 \le \theta \le \pi} s^{\rm in}(\theta) = \min_{0 \le y \le 1} s^{\rm out}(y) \,,
\end{equation}
where $s^{\rm in}(\theta)$ and $s^{\rm out}(y)$ is the invariant mass of the incoming and outgoing particles, respectively, given by the symbolic reaction equations
\begin{equation}
\gamma_{\rm HE} + \gamma_{\rm bg} \rightarrow e^{+}_{\rm HE} + e^{-}_{\rm HE}
\end{equation}
for pair production, and
\begin{equation}
e^{\pm}_{\rm HE} + \gamma_{\rm bg} \rightarrow e^{\pm} + \gamma_{\rm HE} 
\end{equation}
for inverse Compton scattering, where in both cases $\gamma_{\rm bg}$ denotes a background photon, while $e_{\rm HE}^{\pm}$ and $\gamma_{\rm HE}$ is a \ac{HE} electron/positron and photon, respectively. The maximum on the left side of the equation should be searched regarding the angle $\theta$ between the propagation directions of the two incoming particles. On the other hand, since the directions outgoing particles for very high energies may be regarded as nearly parallel, the minimum on the right hand side should be searched regarding the fraction $y$ of the total energy carried by one of the outgoing particles.

This results in the background photon threshold energy, $\epsilon_{\rm thr}$, given by
\begin{equation}
\epsilon_{\rm thr} =
\begin{cases}
k_{\gamma} \left[\left( \frac{m_{e}}{k_{\gamma}} \right)^{2} + \frac{1}{4}\left( \frac{\chi_{n}^{e}}{2^{n}} - \chi_{n}^{\gamma} \right) \left(\frac{k_{\gamma}}{M_{\rm Pl}}\right)^{n} \right] & \text{for pair production,}
\\
0 & \text{for inverse Compton scattering,}
\end{cases}
\end{equation}
or, equivalently, in the threshold invariant mass, $s_{\rm thr}$, which may be written as
\begin{equation} \label{sthr}
s_{\rm thr} =
\begin{cases}
k_{\gamma}^{2} \left[ 4 \left(\frac{m_{e}}{k_{\gamma}}\right)^{2} + \frac{\chi_{n}^{e}}{2^{n}} \left(\frac{k_{\gamma}}{M_{\rm Pl}}\right)^{n} \right] & \text{for pair production,}
\\
p_{e}^{2} \left[ \left(\frac{m_{e}}{p_{e}}\right)^{2} + \chi_{n}^{e} \left(\frac{p_{e}}{M_{\rm Pl}}\right)^{n} \right] & \text{for inverse Compton scattering.}
\end{cases}
\end{equation}

Using these results, one then can calculate the mean free path $\lambda$ by \cite{Addazi:2021xuf}
\begin{equation} \label{LIVMFP}
\begin{split}
\lambda^{-1} = &\frac{1}{2 p_{\rm in}} \int_{s_{\rm thr}}^{\infty} \frac{n_{\rm bg}\left(\frac{s^{*} - \mathfrak{S}(p_{\rm in})}{4 p_{\rm in}}\right)}{\left[ s^{*} - \mathfrak{S}(p_{\rm in}) \right]^{2}} \int_{\mathfrak{S}(p_{\rm in})}^{s^{*}} \sigma(s) \left[ s - \mathfrak{S}(p_{\rm in})  \right] \, {\rm d}s \, {\rm d}s^{*}\,,
\end{split}
\end{equation}
where $p_{\rm in}$, $m_{\rm in}$ and $\chi_{n}^{\rm in}$ is the momentum, mass and \ac{LIV} parameter of the incoming \ac{VHE} particle, respectively, $n_{\rm bg}(\epsilon)$ is the ambient photon density of the background photon filed, and $\sigma(s)$ is the cross section of the considered reaction, while $s_{\rm thr}$ is given by Eq.~(\ref{sthr}) and $\mathfrak{S}(p_{\rm in})$ by 
\begin{equation}
\mathfrak{S}(p_{\rm in}) \equiv {p_{\rm in}^{2} \left[ \left(\frac{m_{\rm in}}{p_{\rm in}}\right)^{2} + \chi_{n}^{\rm in} \left( \frac{p_{\rm in}}{M_{\rm Pl}}\right)^{n} \right]} \,.
\end{equation}

\begin{figure}[htp] 
  \centering
  \includegraphics[width=0.495\textwidth]{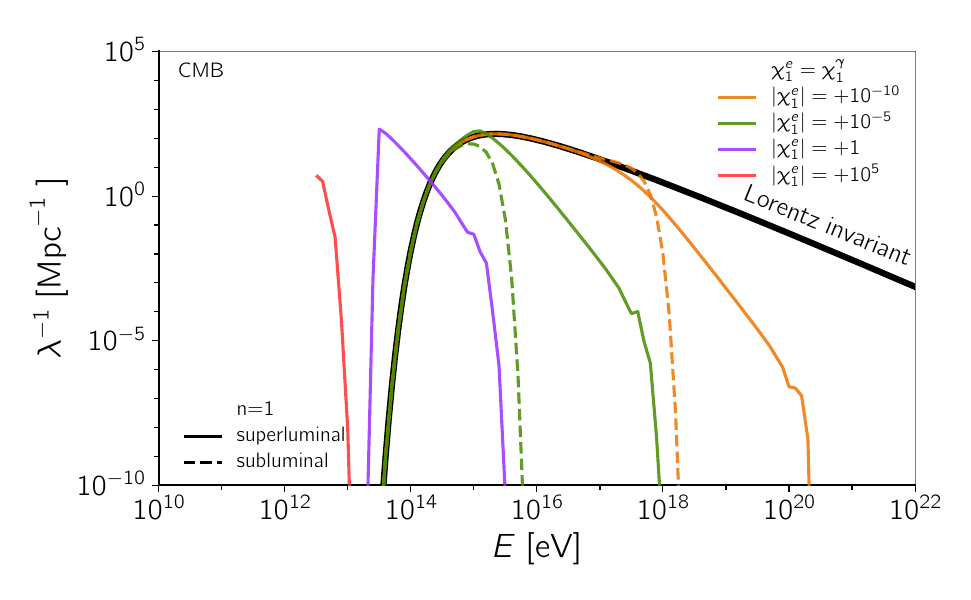}
  \includegraphics[width=0.495\textwidth]{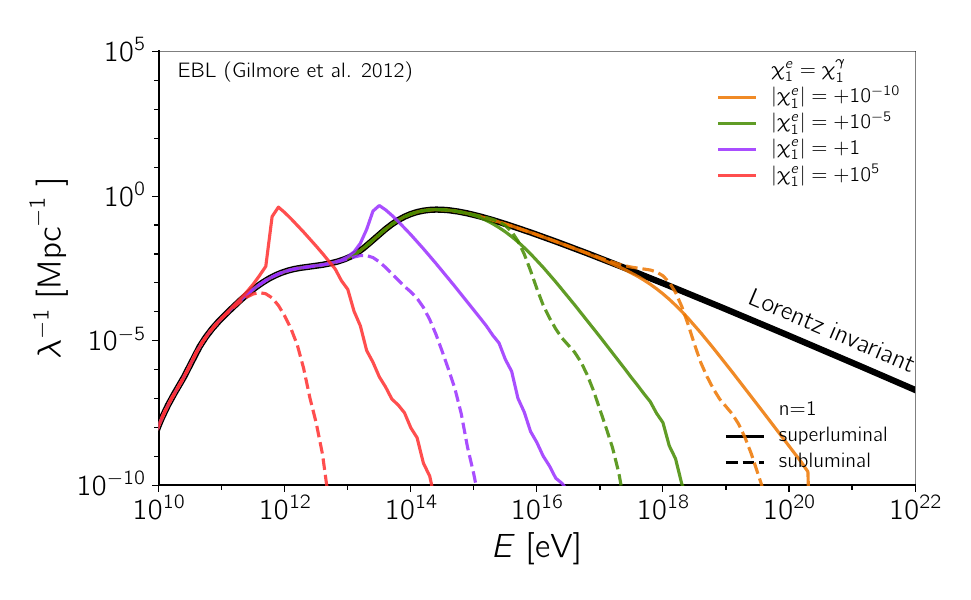}
  \includegraphics[width=0.495\textwidth]{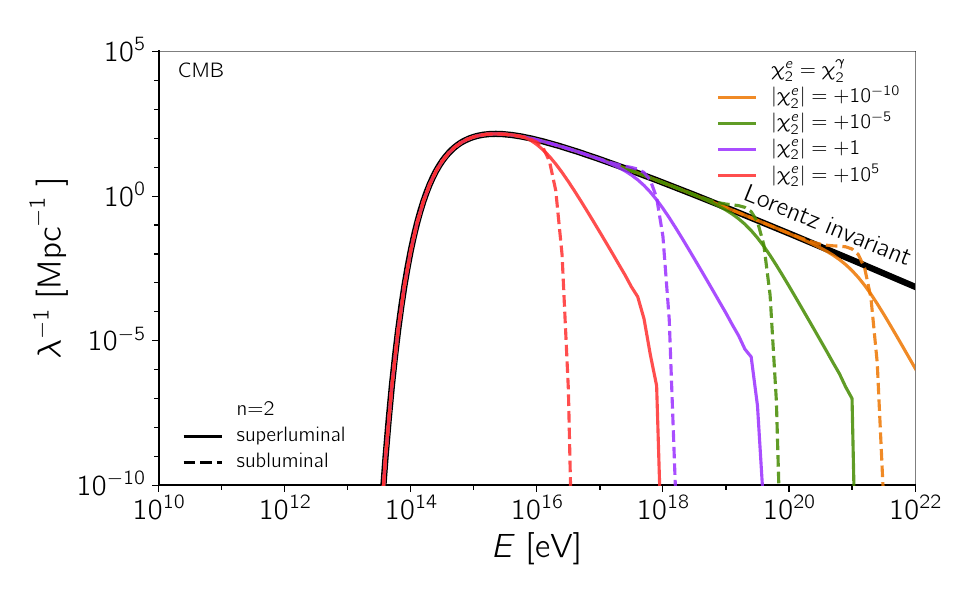}
  \includegraphics[width=0.495\textwidth]{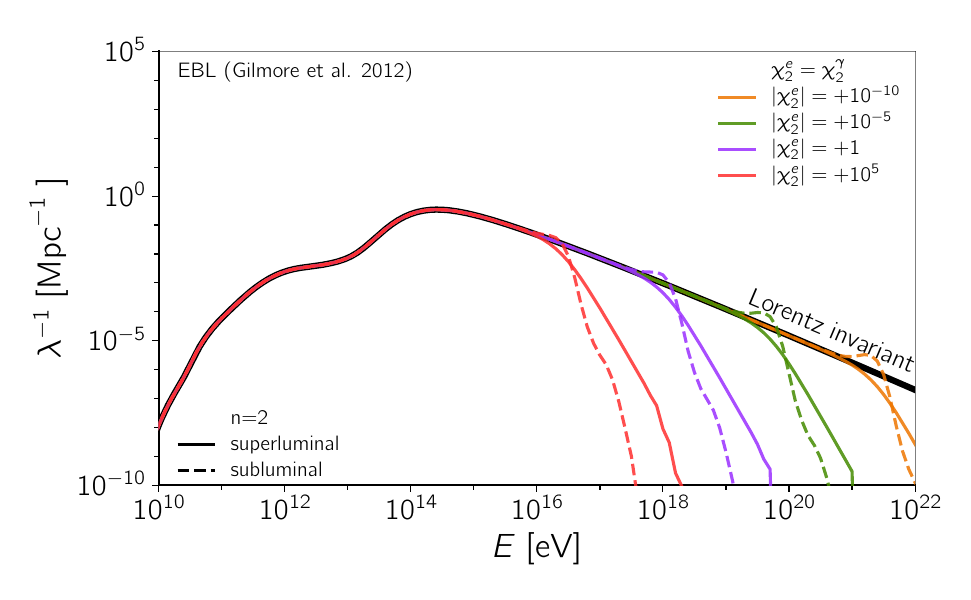}
  \caption{Examples of mean free paths for pair production including \ac{LIV} calculated according to Eq.~(\ref{LIVMFP}). The left panel corresponds to the CMB, whereas the right one corresponds to the EBL model from \cite{Gilmore:2011ks}. The upper panels depict the case of first order LIV, whereas the lower ones are for second order \ac{LIV}. The \ac{LIV} parameters for each case are indicated in the legend.}
  \label{fig:LIVMFP_PP}
\end{figure}

\begin{figure}[htp] 
  \centering
  \includegraphics[width=0.495\textwidth]{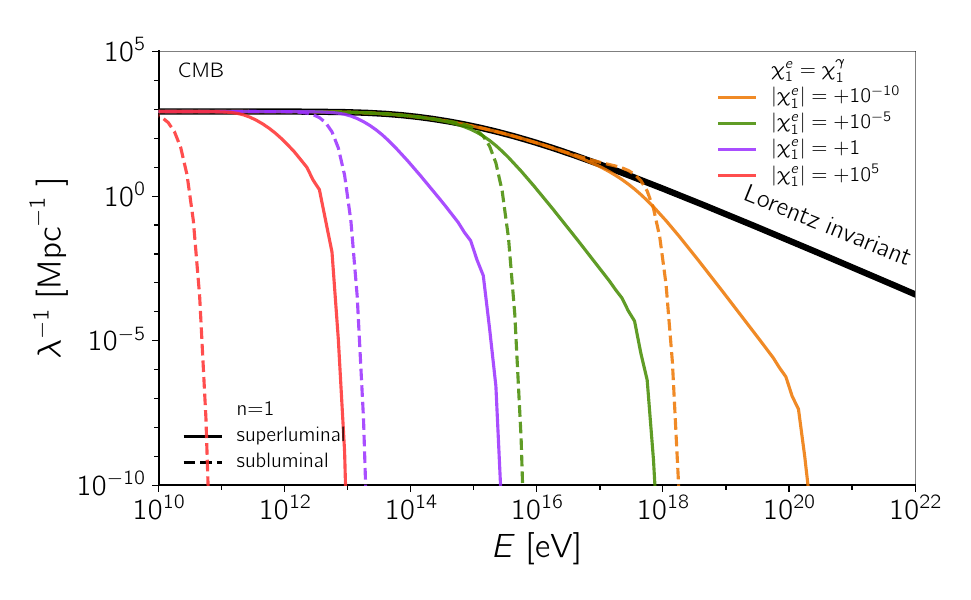}
  \includegraphics[width=0.495\textwidth]{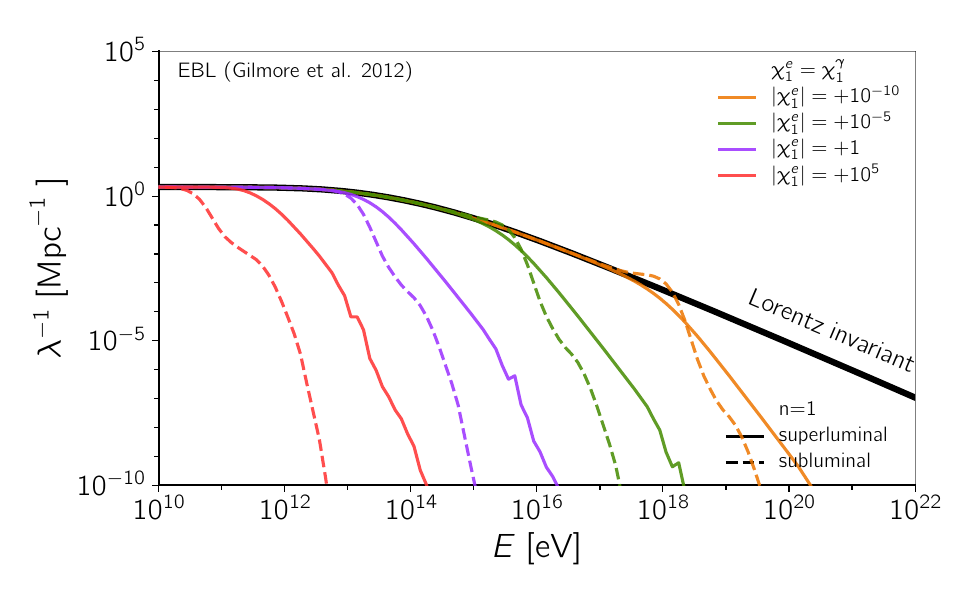}
  \includegraphics[width=0.495\textwidth]{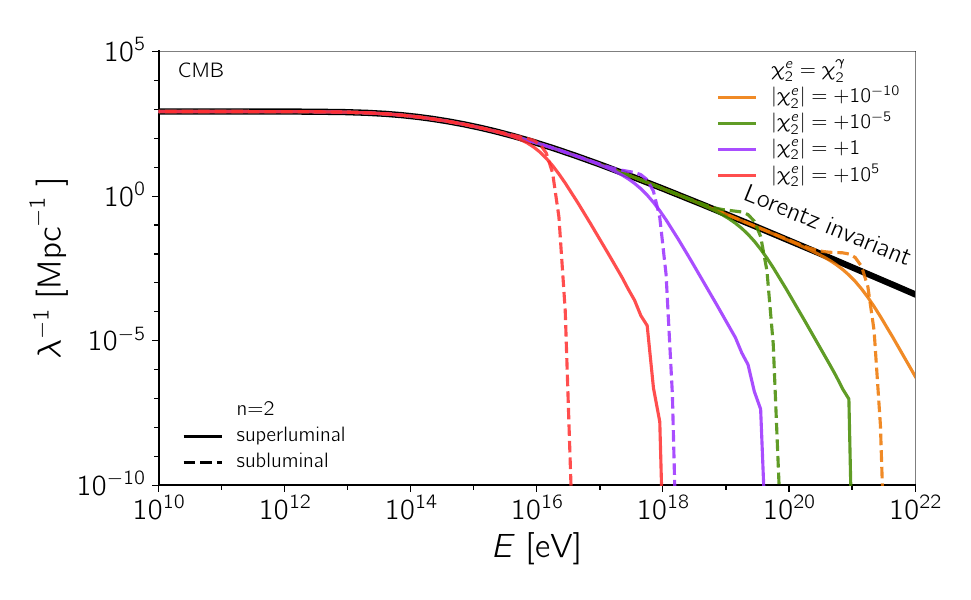}
  \includegraphics[width=0.495\textwidth]{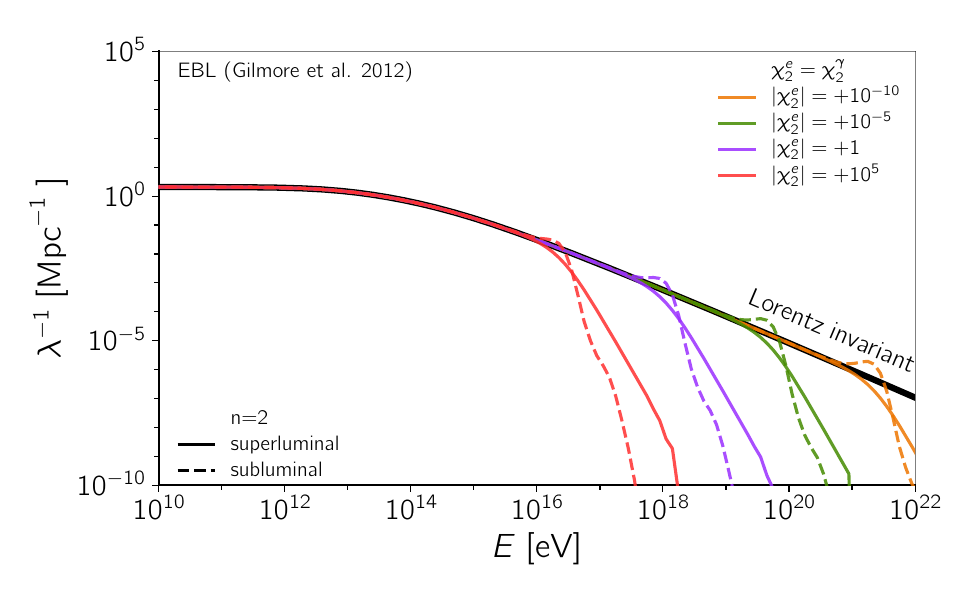}
  \caption{Examples of mean free paths for inverse Compton scattering including \ac{LIV} calculated according to Eq.~(\ref{LIVMFP}). The left panel corresponds to the CMB, whereas the right one corresponds to the EBL model from \cite{Gilmore:2011ks}. The upper panels depict the case of first order LIV, whereas the lower ones are for second order \ac{LIV}. The \ac{LIV} parameters for each case are indicated in the legend.}
  \label{fig:LIVMFP_ICS}
\end{figure}

In Fig.~\ref{fig:LIVMFP_PP} we show some examples of mean free paths for pair production by gamma rays according to Eq.~(\ref{LIVMFP}). Fig.~\ref{fig:LIVMFP_ICS} illustrates the same scenarios considering inverse Compton scattering.

\subsection{New Electromagnetic Processes}\label{ssec:newProcesses}

In addition, we also take into account the most important ``new'' reactions, i.e.~reactions which are only possible when Lorentz invariance is violated: the vacuum Cherenkov effect and photon decay.

The phenomenon known  as \acl{VC} effect can be described by the spontaneous emission of a photon by a massive particle, in this case, electrons and positrons. This is represented by the reaction
\begin{equation}
e^{\pm} \rightarrow e^{\pm} + \gamma,
\end{equation}
Within our framework, we make the assumption that if the parent's energy exceeds the threshold for \ac{VC}, it will emit one or more photons until its energy falls below this threshold. To streamline our approach, we adopt a ``binary'' methodology: if the electron or positron's energy is \textit{above} the threshold, we reduce its energy to match the threshold value. Conversely, if the energy falls \textit{below} the threshold, we do not consider the \ac{VC} effect. This evaluation is conducted at each propagation step. The specific threshold value we utilize is derived from the findings presented in \cite{Jacobson:2002hd}, which we summarize in the following section.

For $n=0$ we get the threshold value
\begin{equation}
p_{\rm thr}^{\rm VC} = 
\begin{cases}
\frac{m_{e}}{\left(\chi_{0}^{e} - \chi_{0}^{\gamma}\right)^{\frac{1}{2}}} & \text{for } \chi_{0}^{e} > \chi_{0}^{\gamma}\,,
\end{cases}
\end{equation}
while for $n=1$ we have
\begin{equation}
p_{\rm thr}^{\rm VC} = 
\begin{cases}
\left( \frac{m_{e}^{2} M_{\rm Pl}}{2 \chi_{1}^{e}} \right)^{\frac{1}{3}} & \text{for }  \chi_{1}^{e} > 0 \text{ and }  \chi_{1}^{\gamma} \ge -3 \chi_{1}^{e} \,, \\
\left[ - \frac{4 m_{e}^{2} M_{\rm Pl} \left(\chi_{1}^{e} + \chi_{1}^{\gamma}\right)}{\left(\chi_{1}^{e} - \chi_{1}^{\gamma}\right)^{2}} \right]^{\frac{1}{3}} & \text{for } \chi_{1}^{\gamma} < \chi_{1}^{e} \le 0 \text{ and }  \chi_{1}^{\gamma} < -3 \chi_{1}^{e} \,.
\end{cases}
\label{eq:pThrVC1}
\end{equation}

The second ``new'' reaction, photon decay (PD), can be described by
\begin{equation}
\gamma \rightarrow e^{+} + e^{-}\,.
\end{equation}
This process involves the \textit{spontaneous} transformation of a high-energy photon into an electron-positron pair without interaction with a background photon. Similarly to the \ac{VC} mechanism, we assume that when the energy of the photon is above a certain threshold for photon decay, it will continually decay into electron-positron pairs until its energy drops below this threshold. Consequently, we again employ a simplified ``binary'' approach: if the photon's energy is \textit{above} the threshold, it is reduced to the threshold value, while if the energy is \textit{below} the threshold, photon decay does not occur. This effectively corresponds to a hard cut-off in the photon spectrum, which is a standard approach in this context in the literature \cite{HAWC:2019gui,Chen:2021hen,LHAASO:2021opi}. We compute the specific threshold value based on the results presented in \cite{Jacobson:2002hd}, which we summarize in the next paragraphs.

\begin{figure}[htb]
    \centering
    \includegraphics[width=0.495\textwidth]{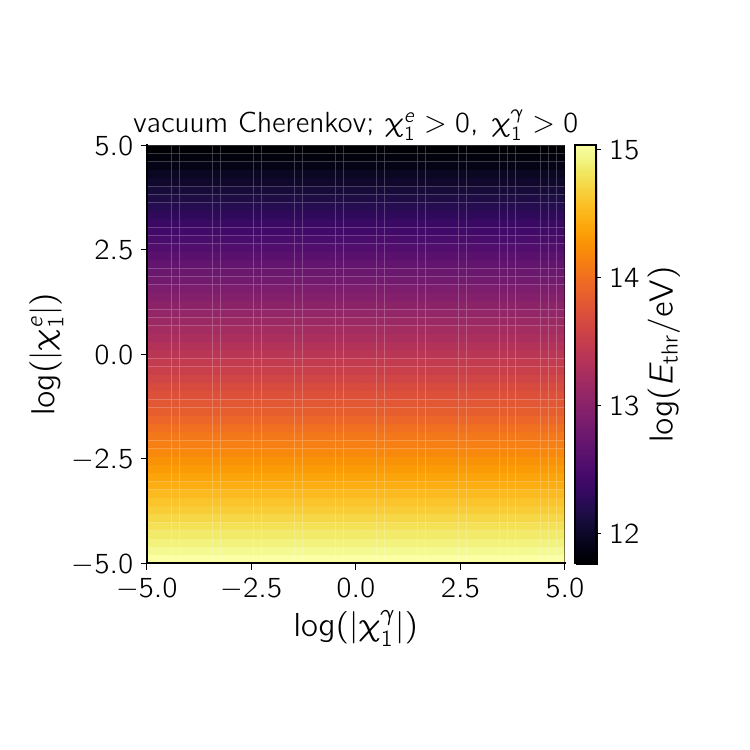}
    \includegraphics[width=0.495\textwidth]{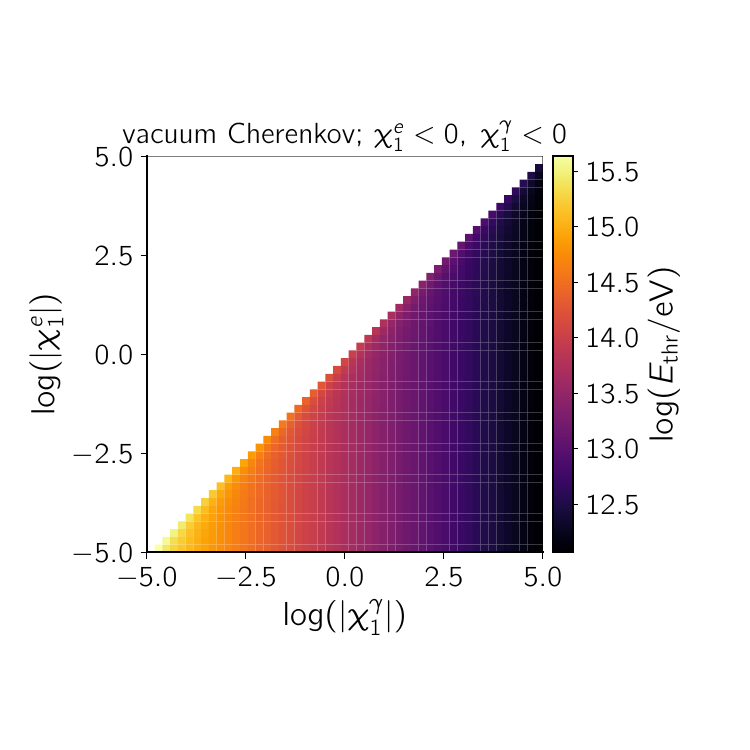}
    \caption{Threshold energy for an electron to emit photons via the vacuum Cherenkov effect, for $n = 1$. The left panel corresponds to the superluminal case ($\chi^\gamma_1 > 0$ and $\chi^e_1 > 0$), whereas the right one depict the subluminal scenario ($\chi^\gamma_1 < 0$ and $\chi^e_1 < 0$). }
    \label{fig:energyThresholds-VC}
\end{figure}

We can now determine the minimum threshold energy that a particle has to have in order for \ac{VC} to take place. This is shown in Fig.~\ref{fig:energyThresholds-VC}. 

It follows from Fig.~\ref{fig:energyThresholds-VC} that, for the superluminal case (left panel), the value of $\chi^e_1$ completely determines the energy threshold of \ac{VC} regardless of $\chi^\gamma_1$. For the subluminal case (right panel) the situation is reversed, with $\left| \chi^e_1 \right|$ exerting little influence on the energy thresholds, which decrease proportionally to $\left| \chi^\gamma_1 \right|$. Note that the subluminal case does not allow for the region of the parameter space left blank, according to Eq.~(\ref{eq:pThrVC1}).

For $n=0$ we get the threshold value
\begin{equation}
k_{\rm thr}^{\rm PD} = 
\begin{cases}
\frac{2 m_{e}}{\left(\chi_{0}^{\gamma} - \chi_{0}^{e}\right)^{\frac{1}{2}}} & \text{for } \chi_{0}^{\gamma} > \chi_{0}^{e}\,,
\end{cases}
\end{equation}
while for $n=1$ we have
\begin{equation}
k_{\rm thr}^{\rm PD} = 
\begin{cases}
\left( \frac{8 m_{e}^{2} M_{\rm Pl}}{2 \chi_{1}^{\gamma} - \chi_{1}^{e}} \right)^{\frac{1}{3}} & \text{for }  \chi_{1}^{\gamma} \ge 0 \,, \\
\left[ - \frac{8 m_{e}^{2} M_{\rm Pl} \chi_{1}^{e}}{\left(\chi_{1}^{e} - \chi_{1}^{\gamma}\right)^{2}} \right]^{\frac{1}{3}} & \text{for } \chi_{1}^{e} < \chi_{1}^{\gamma} < 0 \,.
\end{cases}
\label{eq:kThrPD1}
\end{equation}

\begin{figure}[htb]
    \centering
    \includegraphics[width=0.495\textwidth]{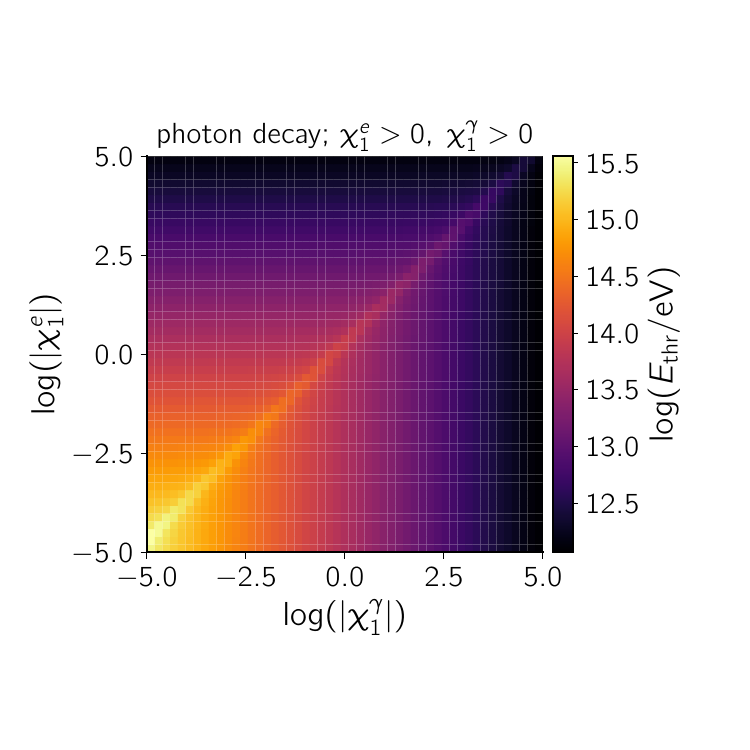}
    \includegraphics[width=0.495\textwidth]{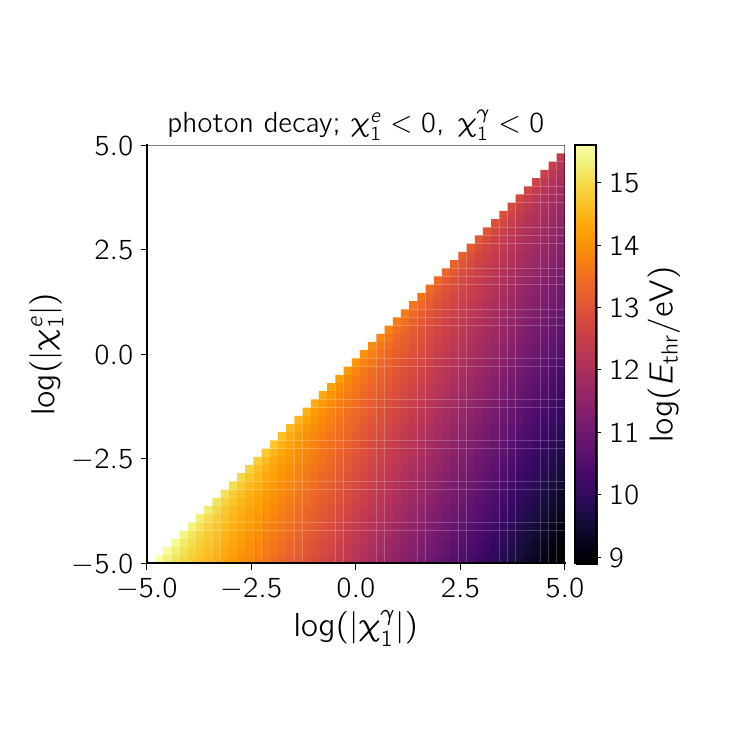}
    \caption{Threshold energy for the photon decay to decay into a pair, for $n = 1$. The left panel corresponds to the superluminal case ($\chi^\gamma _1 > 0$ and $\chi^e _1 > 0$), whereas the right one depict the subluminal scenario ($\chi^\gamma _1 < 0$ and $\chi^e _1 < 0$).}
    \label{fig:energyThresholds-PD}
\end{figure}

The threshold energy for PD is shown in Fig.~\ref{fig:energyThresholds-PD}. 

Fig.~\ref{fig:energyThresholds-PD} indicates that the energy threshold for photon decay in the superluminal case (left panel) decreases as $\chi^e_1$ and $\chi^\gamma_1$ jointly increase. The subluminal case (right panel) has a forbidden region, according to Eq.~(\ref{eq:kThrPD1}), wherein photon decay does not take place. Interestingly, within the allowed region of the parameter space, the energy threshold decreases for large values of $\left| \chi^\gamma_1 \right|$, as expected, but it decreases for lower $\left| \chi^e_1 \right|$. 

\section{Simulations of Electromagnetic Cascades with Lorentz Invariance Violation} \label{sec:Simulations}

To simulate the propagation of gamma-ray--induced electromagnetic cascades, we extend the CRPropa code~\cite{alvesbatista2016a, AlvesBatista:2022vem}, tapping into the already existing structure of this framework to calculate the relevant kinematics in the presence of \ac{LIV}~\cite{Saveliev:2023efw}. 

CRPropa\footnote{\url{https://crpropa.desy.de/}} is an advanced framework for simulating the propagation of high and ultra-high-energy messengers. It implements the main interactions relevant for the propagation of each type of messenger, using a Monte Carlo approach. For this work, we modified the existing modules that handle electromagnetic interactions~\cite{Heiter:2017cev}, which compute the kinematics of pair production and inverse Compton scattering. These modules make use of the routines provided in CRPropa's auxiliary repository~\footnote{\url{https://github.com/CRPropa/CRPropa3-data}}. We also implemented new modules for treating \acl{VC} and \acl{PD}, taking advantage of the code's modular structure.

\bigskip

In the following sections we discuss the phenomenology of \ac{LIV} using simulations obtained with these tools. We select a few scenarios to motivate the discussion, but it is not our goal to compare simulations with observations at this stage. Note that we restrict our analysis to $E < 1 \; \text{PeV}$, considering the substantial computational costs of performing Monte Carlo simulations at higher energies. For this reason, we only present results for the $n=1$ case, since for $n \geq 2$ there is not significant different between the Lorentz invariant and the \ac{LIV} scenarios in the energy range to which we are limited.

We start off by considering the case of \ac{LIV} affecting the propagation of photons (Sec.~\ref{ssec:sim_photons}). Then, in Sec.~\ref{ssec:sim_electrons}, we discuss the case of electrons. In Sec.~\ref{ssec:sim_all} we provide comprehensive simulations comprising all effects considered, namely \ac{PP}, \ac{ICS}, PD, and \ac{VC}. In Sec.~\ref{ssec:sim_spec} we examine the effect of intrinsic source properties on gamma-ray fluxes within the context of \ac{LIV}.

\subsection{Simulations: Photon Propagation}\label{ssec:sim_photons}

We first consider the propagation of photons only, disregarding any electrons that might have been produced. For the sake of discussion, we first consider a benchmark scenario for two astrophysical sources located at $z=0.03$ and $z=0.14$, corresponding to the blazars Mrk~421 and 1ES~0229+200, respectively. The emission spectrum is assumed to be a power-law of the form $E^{-2}$, with a sharp cut-off at $1 \; \text{PeV}$. While this is not necessarily a realistic scenario due to exceedingly high cut-off energy, it does allow us to discuss the main features.

\begin{figure}[htb!]
    \centering
    \includegraphics[width=0.495\textwidth]{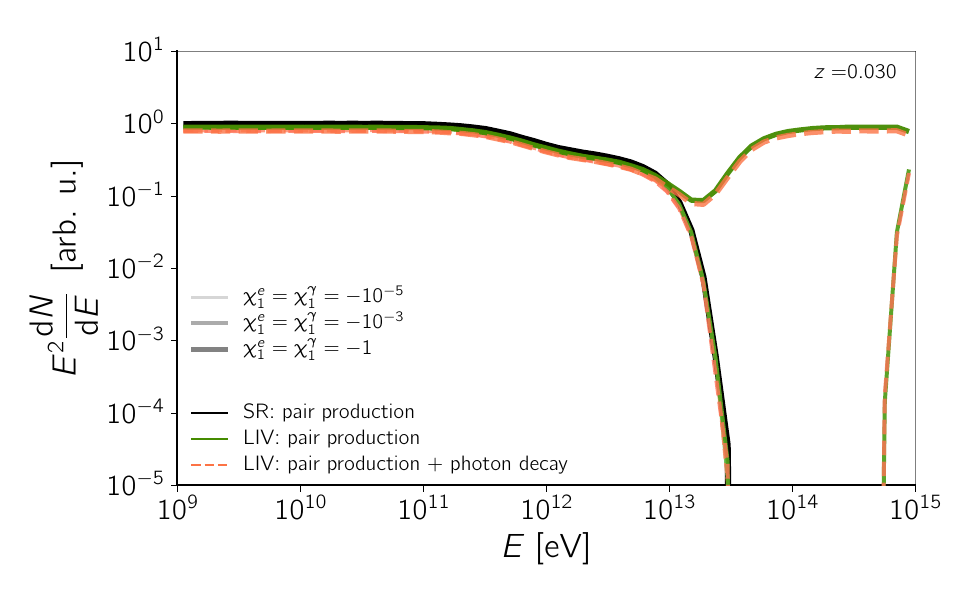}
    \includegraphics[width=0.495\textwidth]{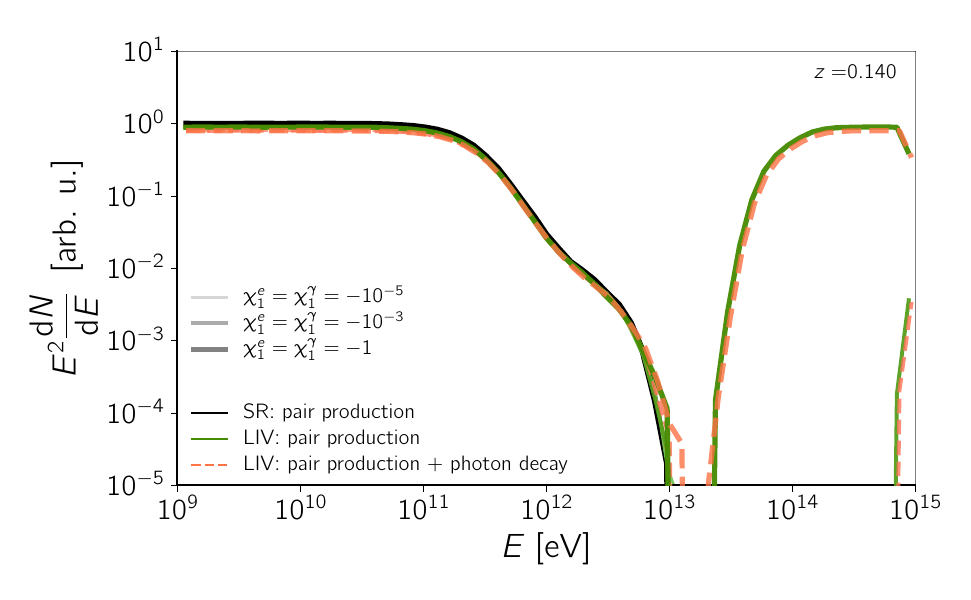}
    \caption{Gamma-ray flux measured at Earth for various scenarios of \ac{LIV} (coloured lines), for $n=1$, considering pair production and photon decay. The special-relativistic (\ac{SR}) scenario is represented by a thick dash-dotted black line. Solid orange lines represent a \ac{LIV} scenario with \ac{PP} only, whereas green lines refer to \ac{PP} and \ac{PD}. Colour gradients and line widths indicate absolute values of \ac{LIV} coefficients, from thin and lighter (smaller coefficients) to thicker and darker shades (larger coefficients).
    }
    \label{fig:specPhotons}
\end{figure}

Due to absorption by background cosmological radiation fields, namely the \ac{CMB} and the \ac{EBL}, a strong suppression of the flux measured at Earth is expected due to pair-production interactions, at some energy range, depending on the source distance, according to Fig.~\ref{fig:LIVMFP_PP}. For instance, any source more distant than $\sim 1 \; \text{Mpc}$ cannot contribute to the gamma-ray flux measured at Earth around $\sim 100 \; \text{TeV}$. In addition, \ac{LIV} generally allows for photon decay as well. These two effects, combined, affect the spectrum. This is shown in Fig.~\ref{fig:specPhotons}.

It follows immediately from Fig.~\ref{fig:specPhotons} that while in the Lorentz-invariant scenario no gamma-ray fluxes are expected at energies above $\gtrsim 10 \; \text{TeV}$, in the presence of \ac{LIV} this is no longer true. An effective change in the transparency of the universe occurs, due to the modification of the interaction thresholds.

\subsection{Simulations: Electron Propagation}\label{ssec:sim_electrons}

We now delve into the phenomena influencing the propagation of electrons and positrons. Our approach involves conducting simulations wherein these particles are introduced at the Galactic centre and subsequently traced to Earth. We deliberately disregard any generated photons within this simulation, to isolate only the relevant processes. It is important to note that a complete assessment would require the inclusion of additional factors, such as deflections caused by the Galactic magnetic field and synchrotron emission. Nevertheless, this simplified scenario effectively serves its purpose: to visually illustrate the alterations in electron fluxes resulting from the breaking of Lorentz symmetry.

To this end, we take the Galactic centre to be our source. We assume it injects electrons. We allow these electrons to propagate to Earth and interact with background photon fields (\ac{CMB} and \ac{EBL}) via \acl{ICS}. 

\begin{figure}[htb!]
    \centering
    \includegraphics[width=0.495\textwidth]{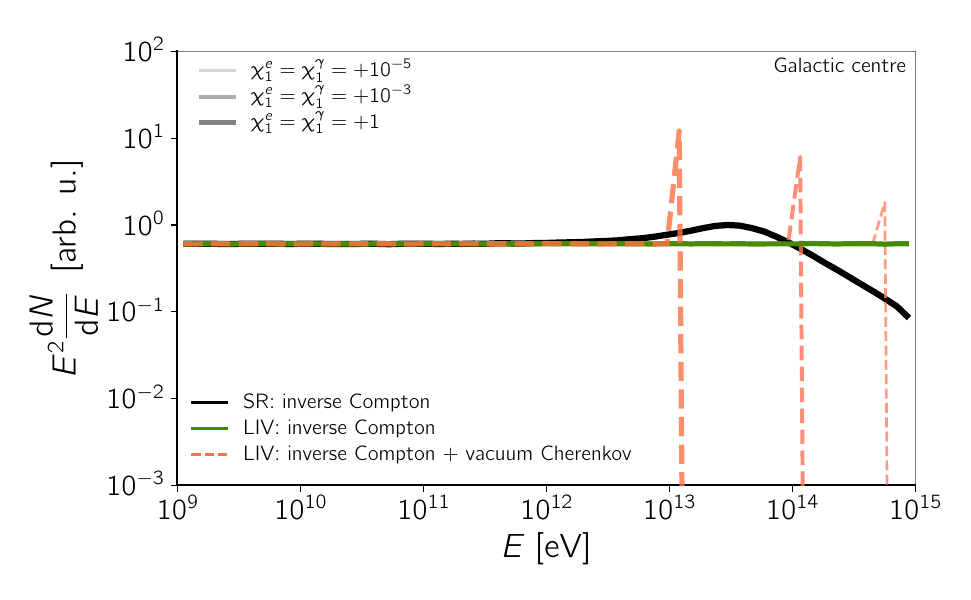}
    \caption{Flux of electrons considering \acl{ICS} and \acl{VC}. The source, in this case, is the Galactic centre. Line styles and colours follow the convention from Fig.~\ref{fig:specPhotons}.}
    \label{fig:specElectrons}
\end{figure}

We proceed to examine the impact of \acl{VC} on the propagation dynamics. To achieve this, we investigate a scenario similar to that for photons, but this time considering \ac{ICS} instead of \ac{PP}, and \ac{VC} in place of \ac{PD}. The results are presented in Fig.~\ref{fig:specElectrons}.

\Acl{VC} is undeniably a crucial ingredient for propagation, leading to a noticeable cut-off in the fluxes, as clearly depicted in Fig.~\ref{fig:specElectrons}. The energy at which this effects becomes apparent is determined by the threshold defined in Eq.~(\ref{eq:pThrVC1}) and illustrated in Fig.~\ref{fig:energyThresholds-VC}. For the \ac{ICS}-only case, there is no suppression due to \ac{VC} emission, as corroborated by the purple curves. \Ac{VC}, on the other hand, lead to pronounced spectral features around the threshold.

\subsection{Simulations: Full LIV Phenomenology}\label{ssec:sim_all}

\begin{figure}[htb!]
    \centering
    \includegraphics[width=0.495\textwidth]{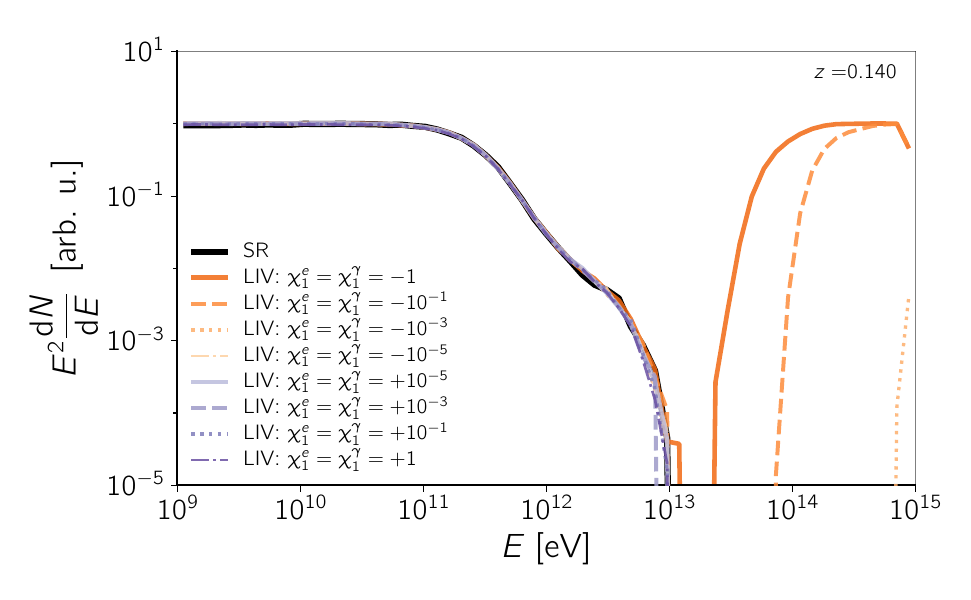}
    \includegraphics[width=0.495\textwidth]{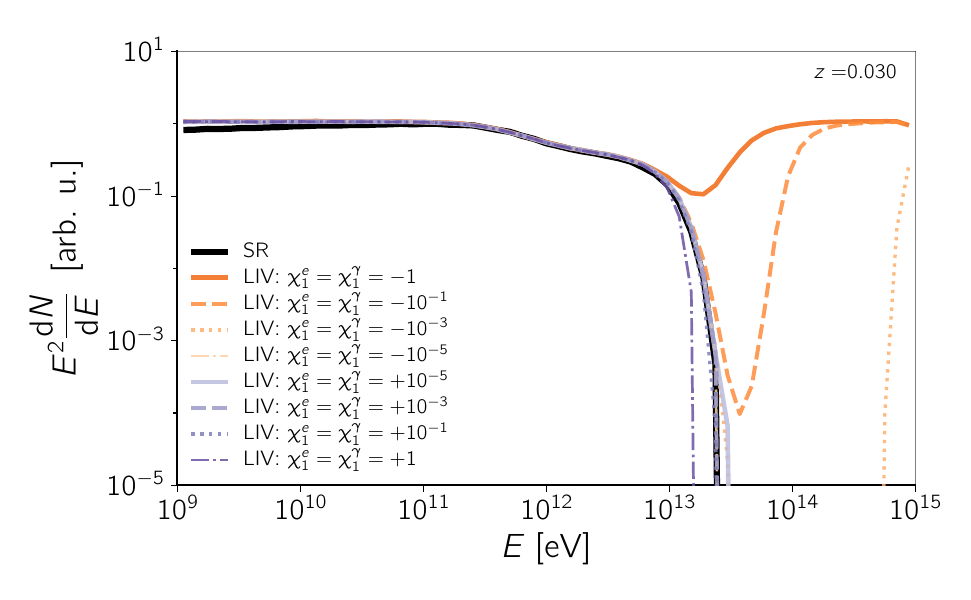}
    \caption{Simulated gamma-ray flux for a source at $z=0.14$ (left panel) and another at $z=0.03$ (right panel), considering \ac{LIV} with $n=1$. Purple lines correspond to positive \ac{LIV} coefficients, i.e., the superluminal case, whereas the orange lines refer to the subluminal case (negative values of $\chi_1$), with the darkness of the shades corresponding to the absolute value of the coefficients. The Lorentz-invariant (\ac{SR}) case is represented by a thick black line.}
    \label{fig:simulationsAll}
\end{figure}

Having isolated each individual effect in the simulations, we now proceed to study a scenario with all relevant phenomenology. We compare the special-relativistic case with the ones including \ac{LIV}. The results are show in Fig.~\ref{fig:simulationsAll}.

The superluminal (purple shades) and subluminal (orange shades) cases of Fig.~\ref{fig:simulationsAll} exhibit very different phenomenological signatures. The latter, in particular, is a clear manifestation of an increased transparency of the universe, thus explaining why a signal is seen at energies above the expected suppression due to the \ac{EBL} absorption of photons.

As expected, a large \ac{LIV} coefficient (darker line shades) enable gamma rays of very high energies to be observed at Earth, since the changes in the kinematical thresholds would render the universe more transparent. Conversely, a small absolute value of the coefficient (lighter line shades) makes the detection of \ac{LIV} signatures more challenging, given its similarity to the Lorentz invariant case.

The case for higher-order \ac{LIV} ($n \geq 2$) leads to much less pronounced features in gamma-ray fluxes. In general, it is hard to separate it from the \ac{SR} scenario. This could, in principle, change if \ac{PD} and \ac{VC} are considered, which we have not done in this in this work.

\subsection{Simulations: Effects of the Intrinsic Properties of the Source}\label{ssec:sim_spec}

The scenarios discussed in the previous sections, while pedagogical, are not realistic. Gamma-ray sources generally emit up to some maximum energy ($E_\text{max}$), determined by an interplay between the maximal energy attainable by their parent cosmic rays (in the case of hadronic sources) and the density of target radiation fields and gas in the surroundings of the emission region. Beyond this energy $E_\text{max}$, the spectrum is suppressed either quickly or more slowly, depending on the exact function describing the suppression. 

\begin{figure}[htb!]
    \includegraphics[width=0.495\textwidth]{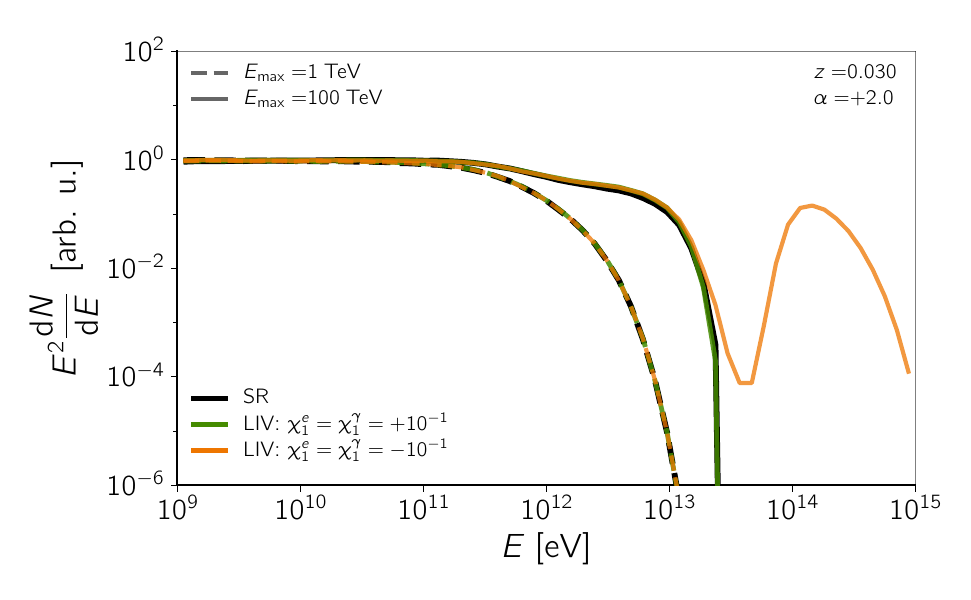}
    \includegraphics[width=0.495\textwidth]{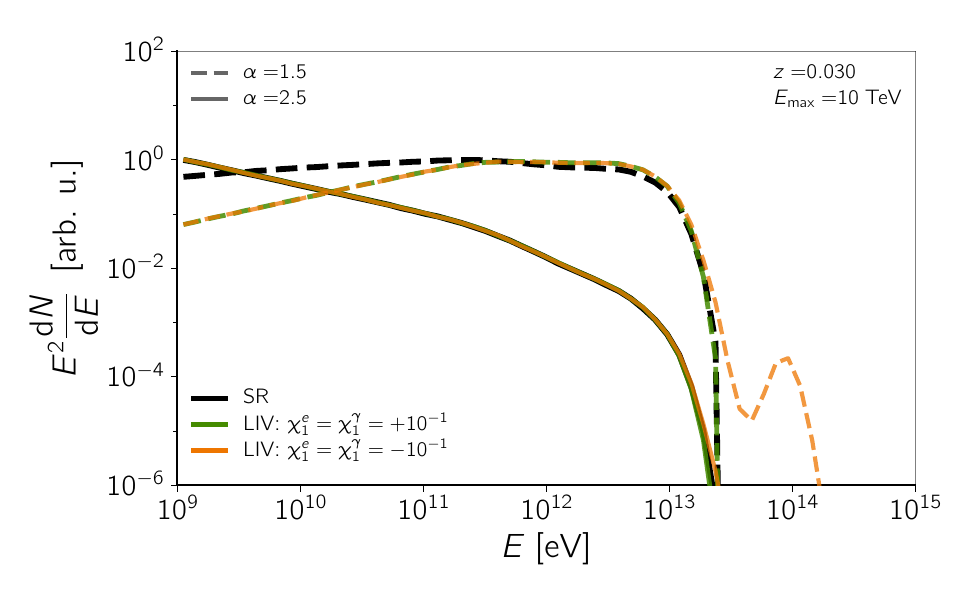}
    \includegraphics[width=0.495\textwidth]{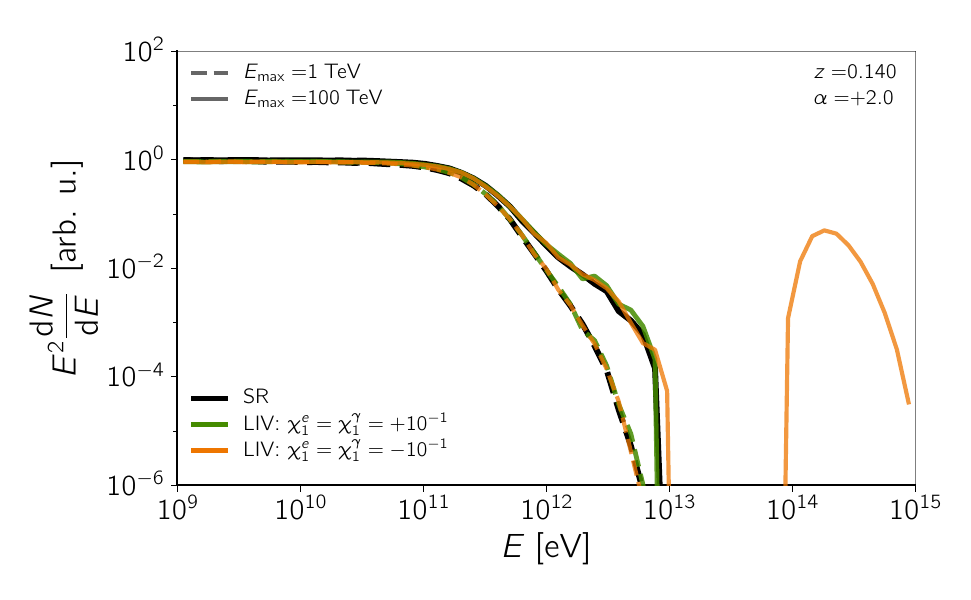}
    \includegraphics[width=0.495\textwidth]{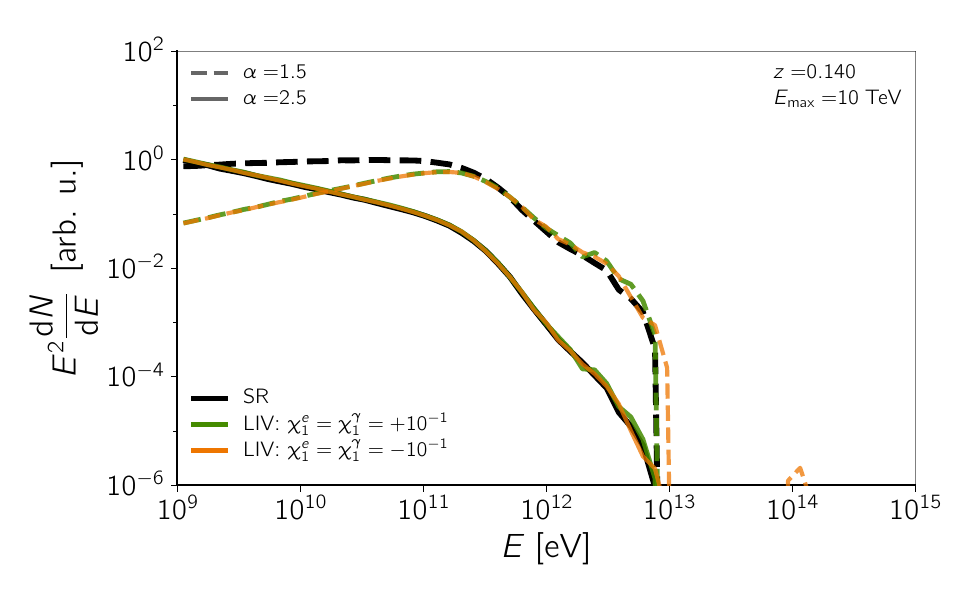}
    \caption{Simulated spectra for two objects, one at $z=0.03$ (upper panels) and the other at $z=0.14$ (lower panels). On the left panels the spectral index is fixed to $\alpha=2$, with the maximal energy ($E_\text{max}$) left to vary. On the right, $E_\text{max}=10 \; \text{TeV}$ while $\alpha$ varies. The Lorentz-invariant case is labelled `SR' and represented by a black line. Green and orange lines refer to \ac{LIV} subluminal and superluminal scenarios, respectively, for \ac{LIV} coefficients $| \chi_1^{e}| = |\chi_1^{\gamma}| = 0.1$.}
    \label{fig:specSrc1}
\end{figure}

\begin{figure}[htb!]
    \includegraphics[width=0.495\textwidth]{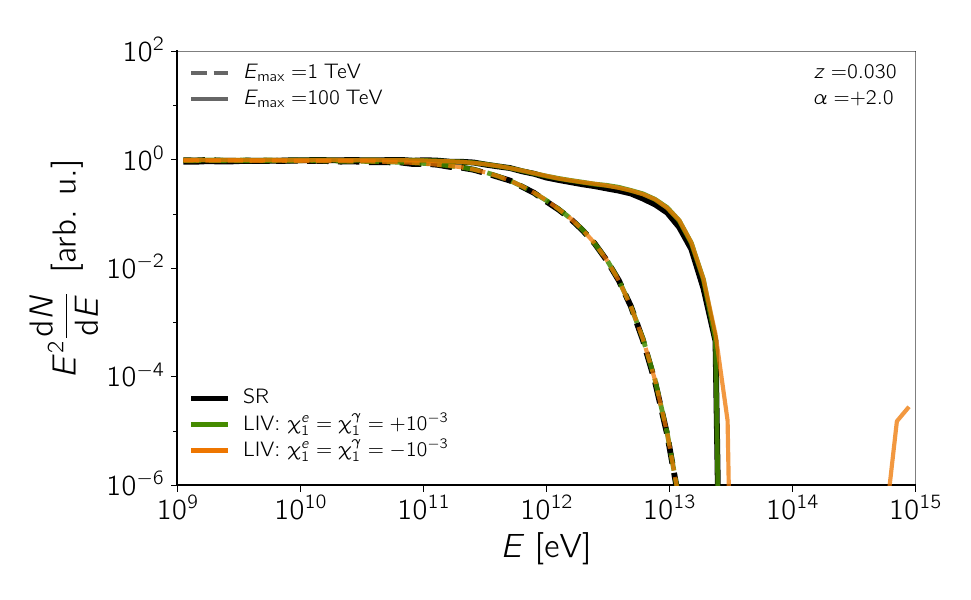}
    \includegraphics[width=0.495\textwidth]{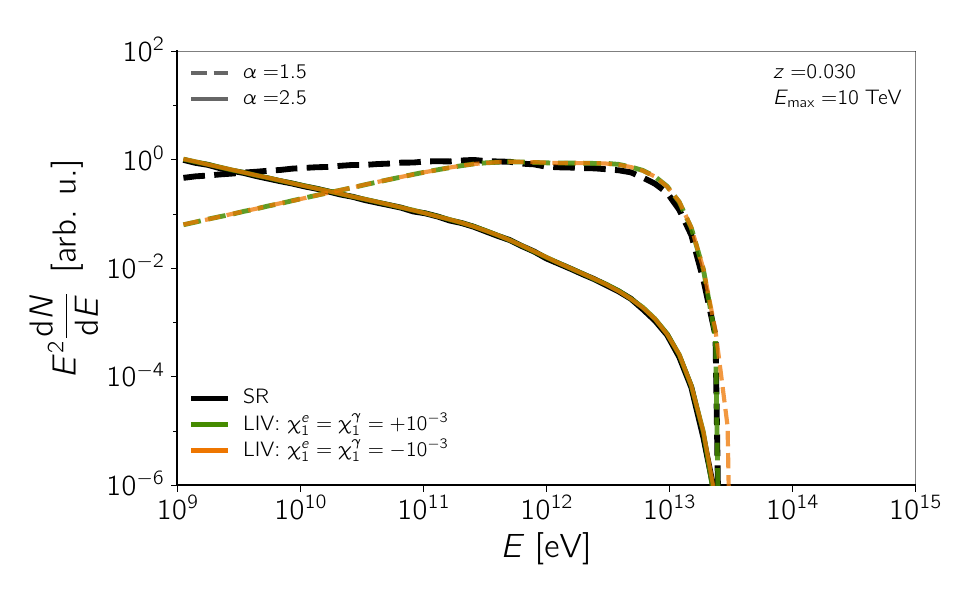}
    \includegraphics[width=0.495\textwidth]{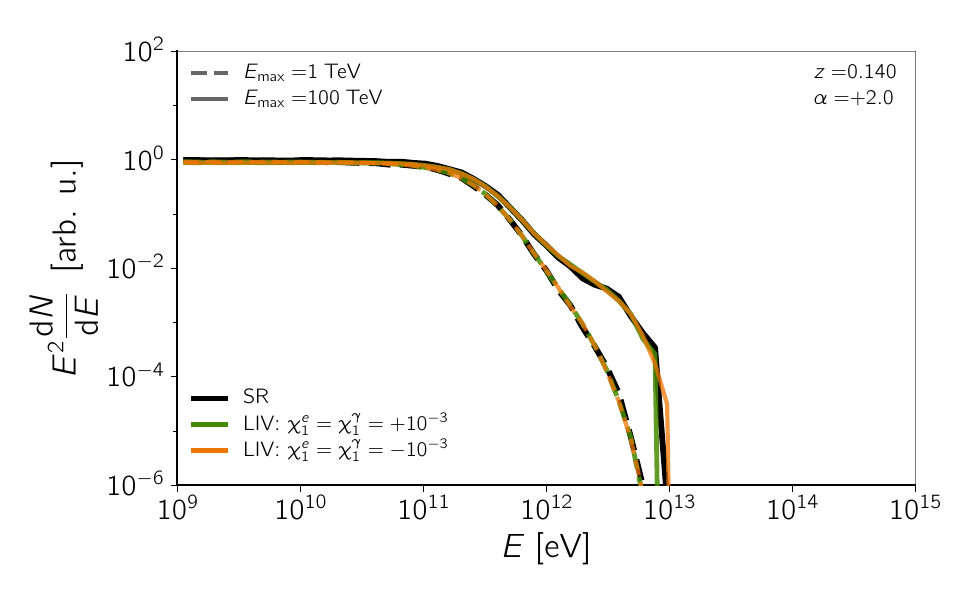}
    \includegraphics[width=0.495\textwidth]{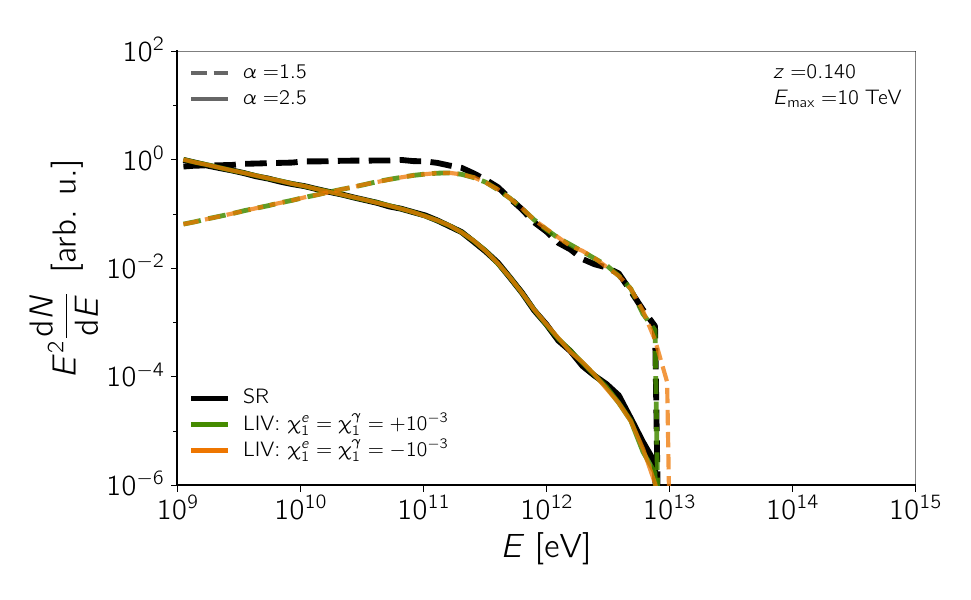}
    \caption{Same as Fig.~\ref{fig:specSrc1} for \ac{LIV} coefficients $| \chi_1^{e}| = |\chi_1^{\gamma}| = 10$.}
    \label{fig:specSrc2}
\end{figure}

To understand the impact of the intrinsic source  spectrum, we assume an emission of the form
\begin{equation}
\dfrac{\text{d}N}{\text{d}E} \propto E^{-\alpha} \exp\left( -\dfrac{E}{E_\text{max}} \right) \,
\end{equation}
where $\alpha$ denotes the spectral index.
We analyse a couple of cases for the same two sources used in Sec.~\ref{ssec:sim_photons}. The results are shown in figures~\ref{fig:specSrc1} and \ref{fig:specSrc2}.

It is evident from Fig.~\ref{fig:specSrc1} that for such low values of the \ac{LIV} coefficient, it is rather challenging to detect any substantial alterations within the same energy range as the \ac{SR} case. The only significant change is a higher-energy bump at $E \gtrsim 100 \; \text{TeV}$ for the subluminal case. Nevertheless, this is only visible for larger values of $E_\text{max}$, as supported by the panels on the left. Similarly, lower spectral indices also favour the detectability of \ac{LIV} effects for the cases studied. If the \ac{LIV} effect is stronger, as illustrated in fig~\ref{fig:specSrc2}, \ac{LIV} signatures are more easily detectable, even at sub-TeV energies. This is particularly  relevant for the harder spectra, shown on the right panels, and it can be partly attributed to the absence of the photon component generated by \acl{ICS} of electrons produced by primary photons that did not interact. In fact, this can be understood in terms of the strong suppression due to \ac{VC}, shown in Fig.~\ref{fig:specElectrons}.

The source distance also plays a role in determining the fluxes to be measured at Earth. While this is not really visible for the smaller \ac{LIV} coefficients from Fig.~\ref{fig:specSrc1}, it can be seen in Fig.~\ref{fig:specSrc2}, by comparing the upper and the lower rows. 

\section{Discussion} \label{sec:Discussion}

It is important to emphasise that the phenomenon of \acl{LIV} extends beyond mere alterations of interaction thresholds, which have dominated the works on the topic so far. Within the confines of a single effective field theory, novel processes that are typically prohibited under a Lorentz-invariant scenario start to surface. Noteworthy examples are photon decay ($\gamma \rightarrow e^{+} + e^{-}$), photon splitting ($\gamma \rightarrow \gamma + \gamma + \gamma$), and vacuum Cherenkov radiation ($e^{\pm} \rightarrow e^{\pm} + \gamma$)~\cite{Jacobson:2002hd, gelmini2005a}. Prior investigations using TeV observations of astrophysical objects, such as the Crab Nebula, have placed constraints on \ac{LIV} in the \ac{QED} sector through searches for these effects \cite{astapov2019a, satunin2019a}.

It should be noted here that the aforementioned ``binary'' treatment of these new reactions (in this paper specifically photon decay and the vacuum Cherenkov effect) is a very simplified approach. Using a more sophisticated treatment, taking into account the full spectrum of the corresponding secondaries, this might  change the overall observed spectrum. However, we chose the simplification, on the one hand, due to the fact that a full treatment involves non-trivial quantum-electrodynamical calculations which are beyond the scope of the phenomenological approach of the present paper. On the other hand, \ac{VC} and \ac{PD} usually occur on length scales much shorter than \ac{ICS} and \ac{PP}, such that the particles undergoing the latter two reactions are already the result of the two former reactions, and therefore the effects may, at least to some degree as a compromise, be treated separately.

One interesting result that is not often discussed in the literature is the crucial role played by the source intrinsic spectrum in determining the detectability of \ac{LIV} effects. Our results indicate that either a large maximal energy ($E_\text{max}$) or a harder spectrum (smaller $\alpha$) enhance the detectability of the signals. Astrophysical sources whose bulk integrated energy lies below $\sim \; \text{TeV}$ are not suitable objects for \ac{LIV} studies, specially if their spectrum is soft ($\alpha \gtrsim 2$). This is not surprising, given that a pronounced cascade contribution is required for interactions to occur and thus to enable \ac{LIV} studies. Several types of objects satisfy these criteria, with certain types of blazars and \acp{GRB} arguably being the most relevant ones~\cite{Bertolami:1999da,Ellis:2011ek,Vasileiou:2013vra,Lang:2018yog}. Extreme blazars, in particular, tend to have a harder spectrum and often their emission is detected up to multi-TeV energies. Nevertheless, it is important to note that intrinsic source properties also depend on propagation details, like intervening magnetic fields, as shown in~\cite{saveliev2021a}.

It can be seen here that several cases presented above contain some kind of flux suppression, which might overlap with the flux suppression due to the interaction of cascade electrons with magnetic fields \cite{Batista:2021rgm} (or with the intergalactic medium \cite{AlvesBatista:2019ipr}). In fact, the other two phenomenological consequences of magnetic fields onto the electromagnetic cascade, namely the time delay and the seeming broadening of a point source, can also result from \ac{LIV}, making it difficult to distinguish the two effects. In particular, this means that if one takes \Ac{LIV} into account, the limits previously derived for intergalactic magnetic fields might me altered.

Several of the cases presented in this work result in a different kind of spectral feature, namely a distinct peak at high energies, which is separated from the rest of the spectrum by a gap (see, for example, the bottom left panel of Fig.~\ref{fig:specSrc1}). If such a feature is observed, one might be tempted to conclude that the two separated parts of the spectrum stem from two different processes/regions at the source emitting particles in two different energy bands or an additional radiation field absorbing photons. Our findings, however, allow for the possibility that it is simply a propagation effect, just with LIV included.

Other factors would affect the results presented. For instance, the \ac{EBL} model determines the interaction rates of pair production (see Fig.~\ref{fig:LIVMFP_PP}). The effect of the \ac{EBL} in the presence of \ac{LIV} would not qualitatevely differ from the Lorentz-invariant case.

\Acp{IGMF} could affect the charged component of the cascade, potentially altering the energy spectra expected at Earth~\cite{Batista:2021rgm}. In fact, this could greatly impact gamma-ray--based estimates of \acp{IGMF} (e.g., ~\cite{Neronov02042010, 2010MNRAS.406L..70T, Finke:2013bua, Biteau:2018tmv, alvesbatista2020a}). Similarly, the interplay between the geometry of emission and \acp{IGMF} could change spectral and angular signatures of gamma rays from misaligned jetted sources~\cite{2041-8205-719-2-L130}.

We have performed this investigation only for electromagnetic interactions, but analogous signatures are expected for other types of travelling messengers, like \acp{UHECR}. In particular, the GZK effect due to the interaction of energetic protons with the \ac{CMB}, if it occurs, could be substantially altered in the presence of \ac{LIV}, as studied in \cite{Scully:2008jp,Maccione:2009ju,Stecker:2009hj,Torri:2020fao}. 

\section{Conclusions and Outlook} \label{sec:Conclusions}

In this study, we have conducted detailed Monte Carlo simulations to investigate the impact of \acl{LIV} on gamma-ray propagation. In addition to the conventional propagation effects, such as \acl{PP} and \acl{ICS}, we have also incorporated two phenomena stemming from the breaking of Lorentz symmetry: \acl{PD} and \acl{VC}. To the best of our knowledge, this work represents the first and most detailed attempt to model the development of electromagnetic cascades in the intergalactic medium in the presence of \ac{LIV}.

Our findings underscore the complex interplay among several types of effects, revealing that, combined, they can lead to strong deviations in the expected flux of extraterrestrial gamma rays reaching Earth. The main phenomenological consequences include a change in the transparency of the universe to high-energy gamma rays, as well as characteristic spectral features that impact the measured fluxes. 

One of our most remarkable findings highlights the role of the often-overlooked \acl{ICS}, which can exert a substantial influence on gamma-ray fluxes in some cases. Incorporating \ac{ICS} into most analyses is essential due to its potential to affect fluxes, especially in the GeV--TeV band. This is a consequence from the intricate interplay between excess (or deficit) of photons at high energies, relative to the predictions of \ac{SR}, and its impact on the generation of secondary photons through \ac{ICS}. Although the effect may be modest in magnitude, it evidently compromises the interpretation of observations and ought to be taken into account, even if only to assess its (in)significance.

Also striking is the fact that a combined consideration of \ac{ICS} and \ac{VC} could, in certain scenarios, lead to striking signatures at \emph{lower energies}.

In the future, we envision expanding our analysis in several directions. One avenue of exploration involves refining our treatments of \acl{PD} and \acl{VC}, which currently are done only for \ac{LIV} of order $n=0$ and $n=1$, in particular by using a complete quantumelectrodynamic treatment, such as, for example, the one presented in \cite{Rubtsov:2012kb} for $n=2$. We also intend to generalise the \ac{LIV} formalism to allow superposition of order (multiple values of $n$ acting together).  Finally, we intend to include effects of magnetic fields to understand how they can affect the charged leptonic component of electromagnetic cascades. 

Furthermore, we also plan to extend our analysis to a proper treatment of the secondaries resulting from the new reactions made possible by \Ac{LIV}. As described in Sec.~\ref{ssec:newProcesses}, in this work we used a very simplified binary approach by just reducing the energy and momentum of the corresponding particle above the threshold to the threshold value. However, preliminary results of further investigations show that the emission of secondaries in both cases is a process with complicated kinematics which means that the energy distribution between them might significantly deviate from the Lorentz-invariant case, possibly compromising the co-linear approximation employed, wherein the momenta of highly relativistic secondaries are parallel to the parent's momentum. Depending on the value of the \Ac{LIV} parameter, this can therefore result in significant observational features.

\section*{Data Availability Statement}

The code used for the simulations in this manuscript is being prepared to be made public. In the meantime, it can be made available upon reasonable request. The data that support the findings of this study are available upon reasonable request from the authors.

\section*{Acknowledgements}

The work of AS is supported by the Russian Science Foundation under grant no.~22-11-00063. RAB is funded by the ``la Caixa'' Foundation (ID 100010434) and the European Union's Horizon~2020 research and innovation program under the Marie Skłodowska-Curie grant agreement No~847648, fellowship code LCF/BQ/PI21/11830030, and by the grants PID2021-125331NB-I00 and CEX2020-001007-S, both funded by MCIN/AEI/10.13039/501100011033 and by ``ERDF A way of making Europe''.

\section*{References}

\providecommand{\newblock}{}

\end{document}